\documentclass[oldversion]{aa} 
\usepackage{graphicx}
\usepackage{txfonts}
\begin{document}
   \title{The Density Variance -- Mach Number Relation in the Taurus Molecular Cloud}

   \titlerunning{The Density Variance -- Mach Number Relation}

   \author{C. M. Brunt
       \inst{}  
          }

   \offprints{C. M. Brunt}

   \institute{Astrophysics Group, School of Physics, University of Exeter, Stocker Road, Exeter, EX4
   4QL, UK\\
              \email{brunt@astro.ex.ac.uk}
             }

   \date{Received ; accepted }


  \abstract
  {Supersonic turbulence in molecular clouds is a key agent in generating density enhancements that may
subsequently go on to form stars. The stronger the turbulence -- the higher the Mach number -- the
more extreme the density fluctuations are expected to be. Numerical models predict an increase in density variance,
$\sigma^{2}_{\rho/\rho_{0}}$, with rms Mach number, $M$ of the form: $\sigma^{2}_{\rho/\rho_{0}} = b^{2}M^{2}$,
where $b$ is a numerically-estimated parameter, and this prediction forms the basis 
of a large number of analytic models of star formation.
We provide an estimate of the parameter $b$ from $^{13}$CO J=1--0 spectral line imaging observations
and extinction mapping of the Taurus molecular cloud, using a recently developed technique that needs information contained
solely in the projected column density field to calculate $\sigma^{2}_{\rho/\rho_{0}}$.
When this is combined with a measurement of the rms Mach number, $M$, we are able to estimate $b$.
We find $b = 0.48^{+0.15}_{-0.11}$, which is consistent with typical numerical estimates, and is characteristic of
turbulent driving that includes a mixture of solenoidal and compressive modes. More conservatively, we
constrain $b$ to lie in the range 0.3--0.8, depending on the influence of sub-resolution structure and the role of
diffuse atomic material in the column density budget (accounting for sub-resolution variance results in higher values
of $b$, while inclusion of more low column density material results in lower values of $b$; the value $b = 0.48$ applies
to material which is predominantly molecular, with no correction for sub-resolution variance). We also report a break
in the Taurus column density power spectrum at a scale of $\sim$~1~pc, and find that the break is associated with 
anisotropy in the power spectrum. The break is observed in both $^{13}$CO and
dust extinction power spectra, which, remarkably, are effectively identical despite detailed spatial differences
between the $^{13}$CO and dust extinction maps. 
}  


   \maketitle
%

\section{Introduction}

Recent years have seen a proliferation of analytical models of star formation that provide prescriptions for star formation
rates and initial mass functions based on physical properties of molecular clouds (Padoan \& Nordlund 2002; Krumholz \& McKee 2005; 
Elmegreen 2008; Hennebelle \& Chabrier 2008; Padoan \& Nordlund 2009; 
Hennebelle \& Chabrier 2009). While these models differ in
their details, they are all fundamentally based on the same increasingly influential idea that has emerged from numerical models:
that the density PDF is lognormal in form for isothermal gas (V\'{a}zquez-Semadeni 1994), with the normalized density 
variance increasing with the rms Mach number ($\sigma^{2}_{\rho/\rho_{0}} = b^{2}M^{2}$ where $\rho$ is the density, $\rho_{0}$ is the
mean density, M is the 3D rms Mach number, and $b$ is a
numerically determined parameter; Padoan, Nordlund, \& Jones 1997).

There is some uncertainty on the value of $b$. Padoan et al (1997) propose $b = 0.5$ in 3D,
while Passot \& V\'{a}zquez-Semadeni (1998) found $b = 1$ using 1D simulations. Federrath, Klessen, \& Schmidt (2008)
suggested that $b = 1/3$ for solenoidal (divergence-free) forcing and $b = 1$ for compressive (curl-free)
forcing in 3D. A value of $b = 0.25$ was recently found by Kritsuk et al (2007) in numerical
simulations that employed a mixture of compressive and solenoidal forcing. Lemaster \& Stone (2008) 
found an non-linear relation between $\sigma^{2}_{\rho/\rho_{0}}$ and $M^{2}$, but it is very similar to
a linear relation with $b = 1/3$ over the range of Mach numbers they analyzed.
They also found that magnetic fields appear to have only a weak effect on the $\sigma^{2}_{\rho/\rho_{0}}$~--~$M^{2}$ 
relation, and noted that the relation may be different in conditions of decaying turbulence.

Currently, observational information on the value of $b$, or the linearity of the $\sigma^{2}_{\rho/\rho_{0}}$~--~$M^{2}$
relation, is very sparse. A major obstacle is the inaccessibility of the 3D density field. Recently, Brunt, 
Federrath, \& Price (2009; BFP) 
have developed and tested a method of calculating $\sigma^{2}_{\rho/\rho_{0}}$ from 
information contained solely in the projected column density field, which we apply to the Taurus molecular
cloud in this paper. Using a technique similar to that of BFP, Padoan, Jones, \& Nordlund (1997) have previously 
estimated a value of $b = 0.5$ ($M = 10$, $\sigma_{\rho/\rho_{0}} = 5$) for the IC~5146 molecular cloud.
In sub-regions of the Perseus molecular cloud Goodman, Pineda, \& Schnee (2009) found no obvious relation between 
Mach number and normalized column density variance, $\sigma^{2}_{N/N_{0}}$. It has been suggested by 
Federrath et al (2008) that this could be due to differing levels of compressive forcing, but it may also be due to 
differing proportions of $\sigma^{2}_{\rho/\rho_{0}}$ projected into $\sigma^{2}_{N/N_{0}}$ (BFP). Alternatively, 
this may imply that there is no obvious relation between $\sigma^{2}_{\rho/\rho_{0}}$ and $M^{2}$.

The aim of this paper is to constrain the $\sigma^{2}_{\rho/\rho_{0}}$~--~$M^{2}$ relation observationally.
In Section~2, we briefly describe the BFP technique, and in Section~3 we apply it to spectral line
imaging observations (Narayanan et al 2008; Goldsmith et al 2008) and dust extinction mapping (Froebrich et al 2007)
of the Taurus molecular cloud to establish an observational estimate of $b$. 
Section~4 provides a summary.

\section{Measuring the 3D Density Variance}

We do not have access to the 3D density field, $\rho(x,y,z)$ in a molecular cloud to measure $\sigma^{2}_{\rho/\rho_{0}}$
directly, but must instead rely on information contained the projected 2D column density field, $N(x,y)$. One can use
Parseval's Theorem to relate the observed variance in the normalized column density field, $\sigma^{2}_{N/N_{0}}$
to the sum of its power spectrum, $P_{N/N_{0}}(k_{x},k_{y})$. It can be shown that $P_{N/N_{0}}(k_{x},k_{y})$
is proportional to the $k_{z} = 0$ cut through the 3D power spectrum, $P_{\rho/\rho_{0}}(k_{x},k_{y},k_{z})$.
Since $\sigma^{2}_{\rho/\rho_{0}}$ can in turn be related, again through Parseval's Theorem, to the sum
of its power spectrum, $P_{\rho/\rho_{0}}$, it is possible, assuming an isotropic density field, to calculate the
ratio, $R = \sigma^{2}_{N/N_{0}} / \sigma^{2}_{\rho/\rho_{0}}$ from measurements made on the column density field
alone. It was shown by BFP that $R$ is given by: 

\begin{eqnarray}
\lefteqn{ R = } \nonumber \\
\lefteqn{\frac{[\displaystyle\sum_{k_{x}=-\lambda/2+1}^{\lambda/2} \sum_{k_{y}=-\lambda/2+1}^{\lambda/2} \langle{P_{N/N_{0}}}\rangle (k) \tilde{B}^{2}(k) ] - P_{N/N_{0}}(0)\tilde{B}^{2}(0)}{[\displaystyle\sum_{k_{x}=-\lambda/2+1}^{\lambda/2} \sum_{k_{y}=-\lambda/2+1}^{\lambda/2} \sum_{k_{z}=-\lambda/2+1}^{\lambda/2} \langle{P_{N/N_{0}}}\rangle (k) \tilde{B}^{2}(k) ] - P_{N/N_{0}}(0)\tilde{B}^{2}(0)} ,}
\label{eqno1}
\end{eqnarray}
where $\lambda$ is the number of pixels along each axis, $\langle{P_{N/N_{0}}}\rangle(k)$ is the azimuthally-averaged power
spectrum of $N/N_{0}$, and $\tilde{B}^{2}(k)$ is the square
of the Fourier space representation of the telescope beam pattern. In equation~\ref{eqno1}, it should be noted that
$\langle{P_{N/N_{0}}}\rangle(k)$ is the power spectrum of the column density field in the absence of beam-smoothing
and instrumental noise.

\begin{figure}
\includegraphics[width=84mm]{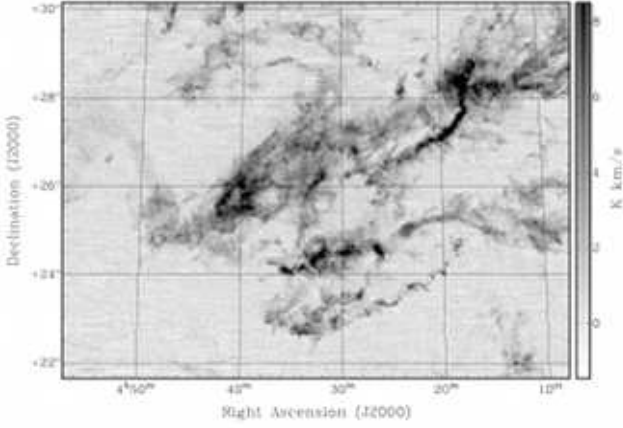}
  \caption{ Integrated intensity map of the $^{13}$CO J=1--0 line over the velocity range $[0,12]$~km~s$^{-1}$ in the Taurus molecular cloud.}
\label{fig:t1}
\end{figure}

An observational measurement of $\sigma^{2}_{N/N_{0}}$ and $R$ can then yield $\sigma^{2}_{\rho/\rho_{0}}$. To
calculate the power spectrum in equation~\ref{eqno1}, zero-padding of the field may be necessary, to reduce
edge discontinuities and to make the field square. 
If a field of size $\lambda_{x} \times \lambda_{y}$ is zero-padded to produce a square field of size 
$\lambda_{px} \times \lambda_{py}$ then $\sigma^{2}_{\rho/\rho_{0}}$ should be calculated via:
\begin{equation}
\sigma^{2}_{\rho/\rho_{0}} = \frac{1}{\eta^{3}} ( 1 + ((\sigma^{2}_{N/N_{0}} + 1)\eta^{2} -1 )/R_{p}) - 1 ,
\label{eqno2}
\end{equation}
where $\eta = (\lambda_{px}\lambda_{py}/\lambda_{x}\lambda_{y})^{\frac{1}{2}}$, and $R_{p}$ is the 2D-to-3D 
variance ratio calculated from the power spectrum of the padded field using equation~\ref{eqno1}.
This assumes that the line-of-sight extent (in pixels) of the density field is $\lambda_{z} = (\lambda_{x}\lambda_{y})^{\frac{1}{2}}$.

The BFP method assumes isotropy in the 3D density field, so equation~(\ref{eqno1}) must be applied
with some caution. 
Using magnetohydrodynamic turbulence simulations, BFP show that the assumption of isotropy is 
valid if the turbulence is super-Alfv\'{e}nic ($M_{A} > 1$) or, failing this, is strongly 
supersonic ($M \gtrsim 10$). The BFP method can be used to derive $\sigma^{2}_{\rho/\rho_{0}}$ to around 10\% accuracy 
if these criteria are met, while up to a factor of 2 uncertainty may be expected in 
the sub-Alfv\'{e}nic, low sonic Mach number regime. 
It should also be noted that the variance calculated at finite resolution is necessarily a lower limit to
the true variance. While this problem can only be addressed by higher resolution observations, estimates
of the expected shortfall in the variance can be made from available information.

\section{Application to the Taurus Molecular Cloud}

\subsection{Measurement of $b$ using $^{13}$CO data only}

We now apply the BFP method to $^{13}$CO (J=1--0) spectral line imaging
observations of the Taurus molecular cloud. Figure~\ref{fig:t1} 
shows the $^{13}$CO emission integrated over the velocity range $[0,12]$~km~s$^{-1}$. In the construction
of this map, we have removed the contribution of the error beam to the observed intensities 
and expressed the resulting intensities on the corrected main beam scale, $T_{MB,c}$ (Bensch, Stutzki, \& Heithausen 2001; 
Brunt et al in prep.; Mottram \& Brunt in prep.) The map is 2048~pixels~$\times$~1529~pixels across, corresponding
to 28~pc~$\times$~21~pc at a distance of 140~pc (Elias 1978).

Following the procedure described in Section~2, we first estimate the variance in the normalized
projected field. For this initial analysis, we use the entire field as represented in Figure~\ref{fig:t1}
with no thresholding of the intensities. We assume that the $^{13}$CO integrated intensity, $I_{13}$, is linearly proportional to 
the column density, $N$. The advantages and disadvantages of this assumption are discussed below in Section~3.2, 
along with an alternative estimate of $b$ using extinction data to calculate the normalized column
density variance. Taking $I_{13} \propto N$, we calculate $\sigma_{N/N_{0}}^{2} = \sigma_{I_{13}/I_{0,13}}^{2}$,
where $I_{0,13}$ is the mean intensity.
The observed variance in the field is the sum of the signal 
variance, $\sigma_{I_{13}}^{2}$, and the noise variance, $\sigma_{noise}^{2}$. We measure $\sigma_{I_{13}}^{2} + \sigma_{noise}^{2} = 2.81$
and $\sigma_{noise}^{2} = 0.42$, giving $\sigma_{I_{13}}^{2} = 2.39$ (units are all (K~km~s$^{-1}$)$^{2}$).
With a measured $I_{0,13} = 1.06$~K~km~s$^{-1}$, we then find $\sigma_{N/N_{0}}^{2} = 2.25$.

\begin{figure}
\includegraphics[width=84mm]{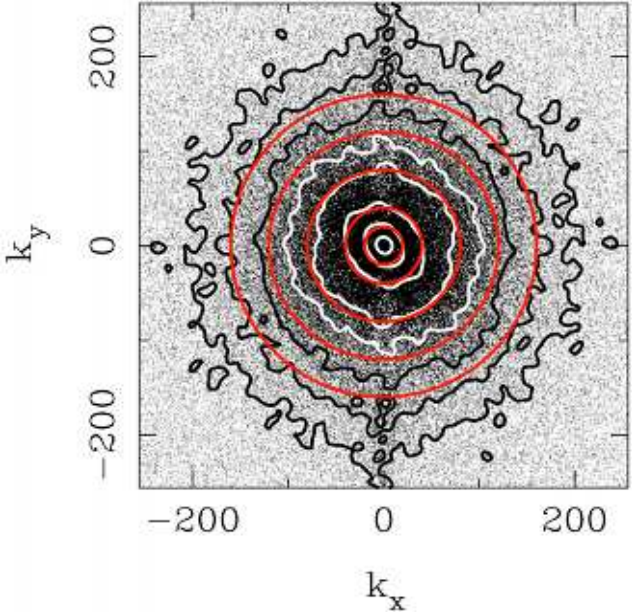}
  \caption{ Central portion of the power spectrum of the Taurus $^{13}$CO J=1--0 integrated intensity map. The black/white contours (smoothed for
clarity) show levels of equal power, while the red circles represent an isotropic power spectrum for reference.}
\label{fig:t2}
\end{figure}

\begin{figure}
\includegraphics[width=84mm]{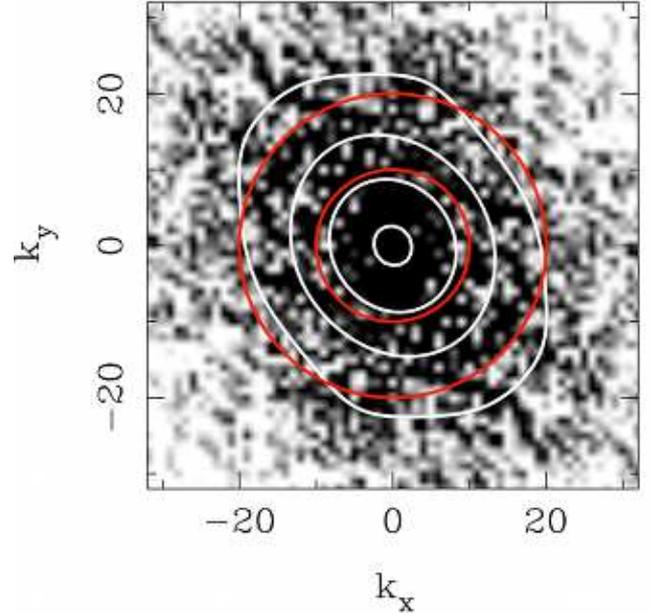}
  \caption{ Zoom in on the central portion of the power spectrum of Figure~\ref{fig:t3}. The white contours (smoothed for
clarity) show levels of equal power, while the red circles represent an isotropic power spectrum for reference.}
\label{fig:t3}
\end{figure}

The power spectrum of the integrated intensity field is now calculated. We use a square field
of size 2048~pixels~$\times$~2048~pixels in which the map is embedded, and compute the power 
spectrum using a Fast Fourier Transform. Application of tapers to smoothly roll-off the 
field edges had an insignificant effect on the result (see Brunt \& Mac Low 2004). We 
calculate the noise floor using the power spectrum of an integrated map made over 
signal-free channels in the data, and this is subtracted from the power spectrum of the 
field. The central portion of the power spectrum is shown in
Figure~\ref{fig:t2} to demonstrate the level of anisotropy present. The black and white contours
(as appropriate to the greyscale level) show levels of equal power, obtained from a smoothed
version of the power spectrum for clarity, and the red circles are
overlayed to show what would be expected for a fully isotropic power spectrum. While Taurus
is often considered ``elongated'' or ``filamentary'', the power spectrum of $^{13}$CO emission
is in fact reasonably isotropic over most spatial frequencies. There is some evidence of
anisotropy at the larger spatial scales (lower $k$), and we further demonstrate this in
Figure~\ref{fig:t3}, which shows a zoom in on the smaller $k$ range of the power spectrum.
Figure~\ref{fig:t3} shows that there is a preferred range in $k$ for anisotropy
to be present, occurring at spatial frequencies near $k = 20$.

\begin{figure}
\includegraphics[width=84mm]{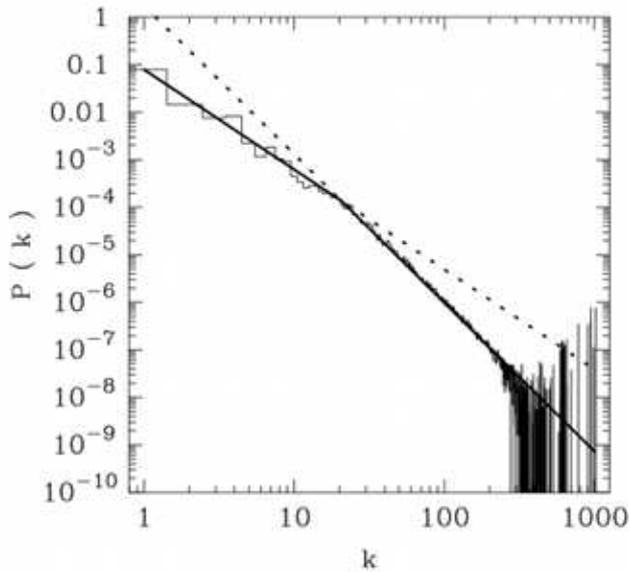}
  \caption{ Angular average of the $^{13}$CO J=1--0 integrated intensity power
 spectrum (histogram) corrected for the noise floor and beam pattern. The solid
lines show the fitted power laws $P(k) \propto k^{-2.1}$  and $P(k) \propto k^{-3.1}$,
for the ranges $k < 21$ and $k > 21$ respectively. The dashed
lines show the continuation of the fitted power laws for reference.}
\label{fig:t4}
\end{figure}

Figure~\ref{fig:t4} shows the angular average of the power spectrum, $P_{N/N_{0}}(k)$, with the
noise floor subtracted and the beam pattern divided out. The heavy lines in Figure~\ref{fig:t4}
display fitted power law forms ($P_{N/N_{0}}(k) \propto k^{-\alpha}$) to two distinct 
$k$-ranges in the power spectrum. The dotted lines continue the fitted forms beyond their 
respective ranges for comparison. For $k < 21$, the best fitting spectral slope is $\alpha = 2.1$, 
while for $k > 21$ we find $\alpha = 3.1$. It is notable that the spectral break
occurs at the preferred scale for anisotropy to be distinguishable in the
power spectrum. The wavelength corresponding to a spatial frequency of $k = 21$ is 
$\sim$~0.5~degrees, or $\sim$~1.2~pc. Previously, Blitz \& Williams (1997), using $^{13}$CO
data but not using the power spectrum, identified a  characteristic scale of $\sim$~0.25--0.5~pc
at which the structure of antenna temperature histograms changes notably. Hartmann (2002) also noted
a characteristic separation scale of $\sim$~0.25~pc in the distribution of young low mass stars in
Taurus. These scales are comparable to about $1/4$ the wavelength of the spectral break in the
power spectrum. The orientation and scale of the anisotropy suggests that it arises from the
repeated filament structure of Taurus, as suggested by Hartmann (2002). 

Figure~\ref{fig:t2} and Figure~\ref{fig:t3} show
that the assumption of isotropy holds reasonably well in 2D, but this is not necessarily
a guarantee of isotropy in 3D, which, nevertheless, we must assume. The presence of a break in the power spectrum
can be accommodated by equation~(\ref{eqno1}) which makes no assumptions, other than isotropy,
on the form of the power spectrum. Evaluating equation~(\ref{eqno1}), we find that $R_{p} = 0.029$,
where the subscript $p$ reflects the fact that padding was employed in the power spectrum calculation.
Evaluating equation~\ref{eqno2} with $\sigma_{N/N_{0}}^{2} = 2.25$ and 
$\eta = (\lambda_{px}\lambda_{py}/\lambda_{x}\lambda_{y})^{\frac{1}{2}} = 1.157$,
we find that $\sigma^{2}_{\rho/\rho_{0}} = 74.23 $. The line-of-sight extent of the field assumed
for this calculation is $\sqrt{28 \times 21}$~pc = 24.25~pc.

Given the high estimated sonic Mach number in Taurus (see below) it is
unlikely that large scale strong anisotropies like those seen in sub-Alfv\'{e}nic conditions 
will be present due to magnetic fields (BFP). Gravitational collapse along magnetic field
lines could in principle generate anisotropy, and indeed the small anisotropy identified in the
power spectrum is oriented as expected in this case (Heyer et al 1987).

\begin{figure}
\includegraphics[width=84mm]{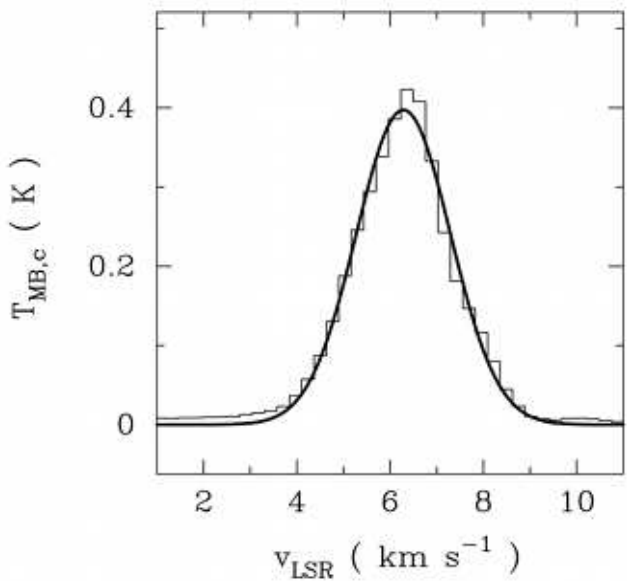}
  \caption{ Global mean spectral line profile for the Taurus $^{13}$CO data (histogram). The heavy solid line is a Gaussian fit, with dispersion
$\sigma_{v_{z}}^{2} = (1.02 \pm 0.06)$~(km~s$^{-1}$)$^{2}$.}
\label{fig:t5}
\end{figure}

To estimate the rms Mach number in Taurus, we now conduct an analysis of the velocity field.
Using the $^{13}$CO spectral line averaged over the field (Figure~\ref{fig:t5}), we measure
the line-of-sight velocity dispersion, $\sigma_{v_{z}}^{2} = (1.02 \pm 0.06)$~(km~s$^{-1}$)$^{2}$ by fitting
a Gaussian profile. The line-of-sight velocity dispersion includes contributions from all internal motions
including those at the largest spatial scales arising from large scale turbulence (Ossenkopf \& Mac Low 2002; 
Brunt 2003; Brunt, Heyer, \& Mac Low 2009).
Assuming isotropy, this implies a 3D velocity dispersion of 
$\sigma_{v}^{2} = 3\sigma_{v_{z}}^{2} = (3.06 \pm 0.18)$~(km~s$^{-1}$)$^{2}$. Taking the Taurus molecular gas to be at a temperature
of 10~K (Goldsmith et al 2008), with a mean molecular weight of 2.72 (Hildebrand 1983), the rms sonic Mach number is $M = 17.6 \pm 1.8$, where the
quoted error estimate is from spectral line fitting and uncertainties in the kinetic temperature (Goldsmith et al (2008) report kinetic 
temperatures for the majority of Taurus between 6 and 12~K; we have taken an uncertainty of $\pm$~2~K). 

Combining these results, our observational estimate of $b$ therefore is $b$~=~$\sigma_{\rho/\rho_{0}}/M$~=~0.49~$\pm$~0.06, where we
have applied an uncertainty of 10\% to the estimated $\sigma^{2}_{\rho/\rho_{0}}$ (BFP 2009). The uncertainty arising
from the assumption of isotropy is difficult to quantify, and further progress on constraining $b$ will require the
analysis of many more molecular clouds. Extension of this analysis to a larger sample of clouds will allow
a test of whether $\sigma^{2}_{\rho/\rho_{0}}$ is indeed proportional to $M^{2}$ and may also
allow investigation of whether $b$ changes due to varying degrees of compressive forcing of the turbulence
(Federrath et al 2008; Federrath et al 2009). The above estimate of the uncertainty in $b$ accounts for measurement
errors on $M$ and known uncertainties in the calculation of $R$ from BFP. The true uncertainty is rather larger than
this, as we discuss in the following sections.

\subsection{Measurement of $b$ using $^{13}$CO and extinction data}

We now consider the consequences of the assumption of linear proportionality between the
$^{13}$CO integrated intensity and the column density. Goodman et al (2009; see also Pineda et al 2008) 
have recently evaluated different methods of measuring column densities in the Perseus molecular clouds. Their findings may
be simply summarised as follows: $^{13}$CO J=1--0 integrated intensity, $I_{13}$, is linearly 
proportional to $A_{V}$ estimated by dust extinction over a limited column density range (over about
a decade or so in $I_{13}$), but is depressed by saturation and/or depletion in the high column density regime; in 
the low column density regime, $I_{13}$ is again depressed by lowered $^{13}$CO abundance and/or subthermal excitation, 
and is insensitive to column densities below some threshold, as evidenced by an offset term in the
$I_{13}$--$A_{V}$ relation. While all these factors impact on the point-to-point
reliability of using $I_{13}$ to derive $N$, their effects on $\sigma^{2}_{I_{13}/I_{0,13}}$
and therefore on the estimated $\sigma^{2}_{N/N_{0}}$ are of more relevance here. 

\begin{figure}
\includegraphics[width=84mm]{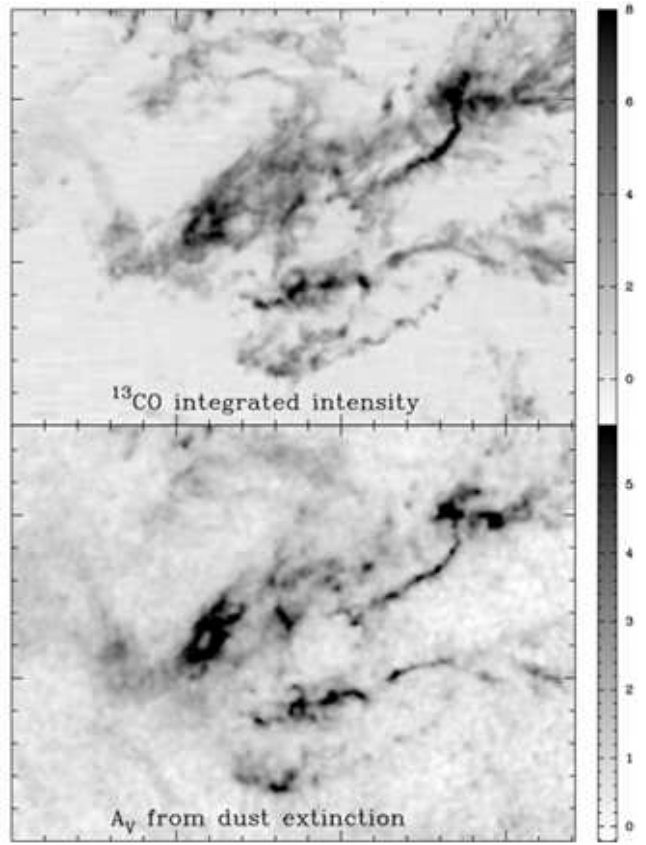}
  \caption{ Comparison of the $^{13}$CO integrated intensity map with the dust extinction map of Froebrich et al (2007).} 
\label{fig:t6}
\end{figure}

To empirically investigate the differences between using dust extinction and $^{13}$CO integrated intensity as a measure
of column density, we re-derived $\sigma^{2}_{N/N_{0}}$ using the dust extinction map generated from 2MASS data by
Froebrich et al (2007). In Figure~\ref{fig:t6} we compare, on the same coordinate grid, the the dust extinction map with the 
$^{13}$CO integrated intensity map convolved to the resolution of the extinction map (4 arcminutes). The greyscale limits of each
field were chosen to represent the fitted relation between the $^{13}$CO integrated intensity, $I_{13}$, and the dust extinction, $A_{V}$,
described below. While the overall spatial distributions of $I_{13}$ and $A_{V}$ are broadly similar, there are detailed differences
in places, and column density traced by $A_{V}$ is present on the periphery of the cloud which is not traced by $I_{13}$. 

\begin{figure}
\includegraphics[width=84mm]{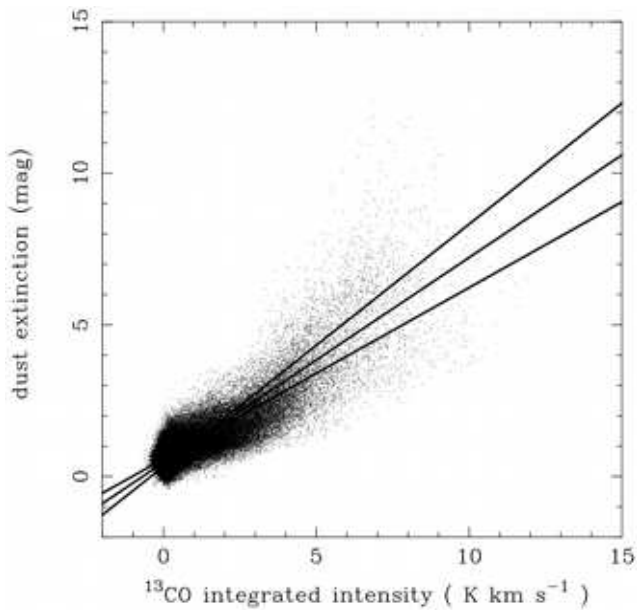}
  \caption{ Plot of dust extinction versus $^{13}$CO integrated intensity, $I_{13}$, for the fields in Figure~\ref{fig:t6}. 
The solid lines are the fitted relations to
$A_{V} = A_{V,D} + C I_{13}$ listed in Table~\ref{tab:1}.}
\label{fig:t7}
\end{figure}

The quantitative relation between $I_{13}$ and $A_{V}$ is shown in Figure~\ref{fig:t7}. 
To these data, we fitted a linear relationship of the form $A_{V} = A_{V,D} + C I_{13}$, where $A_{V,D}$ and $C$ are constants.
We use three different regression methods (Isobe et al 1990) and the fitted parameters, $A_{V,D}$ and $C$, are listed
in Table~\ref{tab:1}, and the fitted lines are overlayed in Figure~\ref{fig:t7}.
While the scatter around these relations is quite large there is only a small number of positions in which 
significant saturation in $I_{13}$ is obvious.

Using the measured $A_{V}$ field to calculate $\sigma^{2}_{N/N_{0}}$, under the assumption that $A_{V}$ is proportional to column density, $N$,
and accounting for a noise variance of 0.04 (Froebrich et al 2007), we find that $\sigma^{2}_{N/N_{0}} = 0.70$. This is significantly
lower than $\sigma^{2}_{N/N_{0}} = 2.25 $ estimated through the $^{13}$CO integrated intensities. There are two contributing
factors to the lower value obtained using $A_{V}$. The first is simply that the fields shown in Figure~\ref{fig:t6} are of
lower angular resolution than the field shown in Figure~\ref{fig:t1}. A calculation of the normalized column density variance using the
lowered resolution $I_{13}$ map results in $\sigma^{2}_{N/N_{0}} = 1.98$, which is lower than that calculated from the higher
resolution field, but still a factor of 2.83 greater than that calculated from the $A_{V}$ field at the same angular resolution.
(Further discussion of the resolution dependence is given below and in Section~3.3).

The second, and dominant, factor contributing to the lower $\sigma^{2}_{N/N_{0}}$ measured from the $A_{V}$ field is the presence of
the offset term, $A_{V,D}$ in the $I_{13} - A_{V}$ relations given above. For reference, note that the addition of a constant positive offset to
$N$ will raise $N_{0}$ but leave $\sigma^{2}_{N}$ unchanged, thereby lowering $\sigma^{2}_{N/N_{0}}$. The true situation is more complicated
that this, since the low column density material is also structured (i.e. contributes variance) --  and it is possible that the column density 
variance scales with column density (Lada et al 1994). It is questionable whether this material should be included in the column density
budget. The excess column density (roughly quantified by $A_{V,D}$) is associated with atomic gas and possibly also diffuse molecular gas ($H_{2}$)
with little or no $^{13}$CO. 

Material in this regime does not contribute to the $^{13}$CO emission, and therefore
also not to the mean spectral line profile, from which the Mach number is estimated. If the material is predominantly
atomic with a possible contribution from diffuse $H_{2}$, it is likely to be physically warmer than the interior regions of the cloud
traced by $^{13}$CO, thereby complicating the assumption of isothermality employed in most of the numerical models {\it and} 
in our calculation of the Mach number. Similarly, $^{13}$CO emission from the subthermally excited (but also likely physically warmer) 
envelope regions will contribute to both to $I_{13}$ and to the mean line profile, but at a reduced 
level in relation to its true column density. 
In addition, while the ``molecular cloud'' part of the extinction is confined to a small region of around 24~pc extent at a distance
of 140~pc, how the ``diffuse'' part of the extinction is generated is rather less well defined, and may include
contributions from all regions along the line-of-sight to Taurus (and beyond). Consequently, the extinction zero level is quite uncertain.
We argue therefore that the selection of material by $^{13}$CO is in fact beneficial to our aims: it preferentially selects the 
interior regions of the molecular cloud, to which the assumption of isothermality is likely to apply, and it limits the
extinction budget to a spatially-confined region at the distance of the cloud. Note also that material contributing 
to $I_{13}$ (and therefore $N$, by assumption) contributes in the same proportion to the mean line profile. 

Inclusion of opacity corrections, using
$^{12}$CO data, also potentially cause as many problems as they solve.
A spatially uniform opacity factor obviously has no impact on the calculated $\sigma^{2}_{N/N_{0}}$ due to the normalization. In
principle, the compression of $I_{13}$ in the high column density regime can be alleviated, but
this effect is likely to be small -- particularly since the strongly-saturated column densities account for a small
fraction of the data. Goodman et al (2009) discuss numerous problems in the $^{12}$CO-based correction for opacity and
excitation in the low column density regime; we also point out here that this is further complicated by the
fact that $^{12}$CO and $^{13}$CO trace different material because of different levels of self-shielding
and radiative trapping. If opacity corrections to $I_{13}$ are made, then for consistency, detailed
(largely impractical) opacity corrections should be made to the line profiles from which the Mach number is calculated,
although for this the point-of-diminishing-returns has certainly long since been reached.

To investigate the effect of the inclusion of the diffuse component, we calculate $\sigma^{2}_{N/N_{0}}$ from the 
field $A_{V}^{'} = A_{V} - A_{V,D}$. For the first of the fitted relations in Table~\ref{tab:1} ($A_{V,D} = 0.579$) we find 
$\sigma^{2}_{N/N_{0}} = 2.73$; from the second ($A_{V,D} = 0.331$) we find $\sigma^{2}_{N/N_{0}} = 1.36$, and from the
third, bisector method ($A_{V,D} = 0.462$) we find $\sigma^{2}_{N/N_{0}} = 1.91$. These values may be summarized
in a combined measurement, with associated uncertainties, of $\sigma^{2}_{N/N_{0}} = 1.91^{+0.82}_{-0.55}$,
which is more compatible with $\sigma^{2}_{N/N_{0}} = 1.98$ calculated from the $^{13}$CO field at the same
angular resolution. This is not surprising, as $A_{V}^{'}$ is, to a good approximation, linearly 
correlated with $I_{13}$, and the value of the coefficient $C$ is irrelevant as we use normalized variances.

\begin{table}
\caption{Fitted parameters to $A_{V} = A_{V,D} + C I_{13}$}             
\label{tab:1}      
\centering   
\renewcommand{\footnoterule}{}  
\begin{tabular}{lccc}
\hline \hline
Regression method & $A_{V,D}$ (mag) & $C$ (mag/(K~km~s$^{-1}$))\\
\hline
$A_{V}$ on $I_{13}$ & 0.579 $\pm$ 0.002  & 0.565 $\pm$ 0.003 \\
$I_{13}$ on $A_{V}$ & 0.331 $\pm$ 0.004  & 0.799 $\pm$ 0.004 \\
Bisector & 0.462 $\pm$ 0.003  & 0.676 $\pm$ 0.003 \\
\hline
\end{tabular}
\end{table}

The dominant contribution to the uncertainty therefore comes from the choice made for the appropriate
treatment of the diffuse component. We argue that the use of $^{13}$CO, or $A_{V}^{'}$, 
to calculate $\sigma^{2}_{N/N_{0}}$ is better-motivated, but we cannot resolve this issue further at present.

\begin{figure}
\includegraphics[width=84mm]{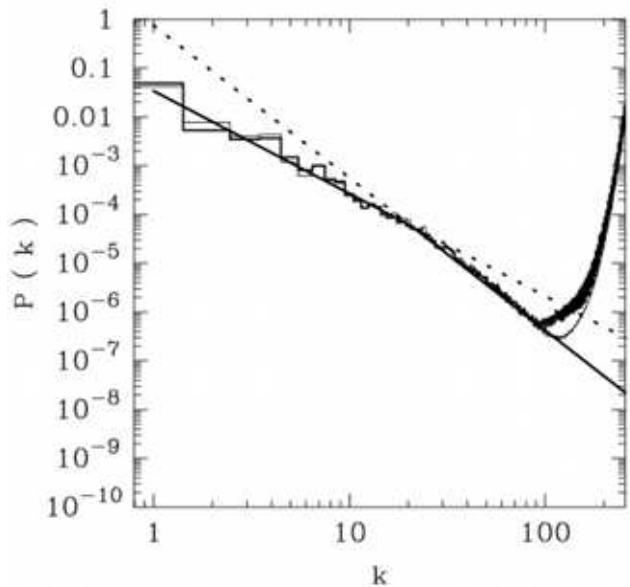}
  \caption{Power spectra of the fields in Figure~\ref{fig:t6}. The light histogram is the power spectrum of $^{13}$CO integrated intensity; the
heavy histogram is the power spectrum of dust extinction. Both fields were scaled to the same global variance of unity before calculating the
power spectra. The power spectra were corrected for the beam pattern but not the noise floor, resulting in an upturn at high wavenumber.
The straight and dashed lines are the fits from Figure~\ref{fig:t4}. }
\label{fig:t8}
\end{figure}

We have noted above that the $\sigma^{2}_{N/N_{0}}$ calculated from $^{13}$CO is, naturally, lower in the
reduced resolution field. Ideally, one should use the highest resolution data available from which to
calculate $\sigma^{2}_{N/N_{0}}$, and this raises the question of how the $A_{V}$-calculated value
of $\sigma^{2}_{N/N_{0}}$ may behave at increased resolution. To investigate this, we calculate the
the power spectra of the $^{13}$CO and $A_{V}$ fields of Figure~\ref{fig:t6}, after scaling each to the same global variance of
unity (the mean of the field is irrelevant). The angular averages of the power spectra
are shown in Figure~\ref{fig:t8}. In these spectra, we have applied ``beam'' corrections, but have not
attempted removal of the noise floor. 

It is remarkable that, despite detailed differences between the two maps, the power spectra are
almost indistinguishable. The fitted power-laws from Figure~\ref{fig:t4} are overlayed for reference -- detailed
correspondance between the two power spectra is notable all the way down to the noise floor(s) near $k \approx 100$.
Comparison of the low resolution $^{13}$CO power spectrum to the fitted slope at high wavenumber from Figure~\ref{fig:t1} 
shows that the noise signature turns on very abruptly, and the power spectrum is very reliable up to this point -- 
this is a useful benchmark to gauge the equivalence of the two low resolution spectra. 
Previously, Padoan et al (2006) have reported differences between $^{13}$CO and extinction power spectra in Taurus, which is
not seen in our analysis. The origin of this discrepancy may lie in the different extinction maps
used by Padoan et al (those of Cambr\'{e}sy 2002), but may also arise from the influence of the
effective ``beam'' size that is not corrected for in their power spectra as the resolution is
varied.

The effective equivalence of the $^{13}$CO and $A_{V}$ power spectra support our suggestion above that it
is in the approximately uniform ``diffuse'' component that the principal statistical differences (at least those
relevant to our analysis) between $^{13}$CO and $A_{V}$ are manifest. This, in turn, suggests that the
$A_{V}$-derived value of $\sigma^{2}_{N/N_{0}}$ that would be obtained at the higher resolution afforded by the $^{13}$CO
data will increase in a similar manner to that seen in the $^{13}$CO -- i.e. an increase from 1.98
to 2.25, or a factor of $\sim$~1.14. Applying this to the $A_{V}^{'}$-based measurements, our estimate 
for $\sigma^{2}_{N/N_{0}}$ at the highest resolution available is $\sigma^{2}_{N/N_{0}} = 2.18^{+0.93}_{-0.63}$.
Converting this to the 3D variance using equation~\ref{eqno2} (with $\eta = 1.157$ and $R_{p} = 0.029$) we
find that $\sigma^{2}_{\rho/\rho_{0}}$ = 72.1$^{+27.8}_{-18.7}$. 
Taking into account the measured Mach number, $M = 17.6 \pm 1.8$, and applying an additional 10\% uncertainty
on the 3D variance arising from the calculation of $R$, we arrive at a 
measured $b = 0.48^{+0.15}_{-0.11}$. (For reference, if $\sigma^{2}_{N/N_{0}}$ is calculated from
$A_{V}$ without subtraction of the diffuse component, we find $b = 0.32$.)

\subsection{The effects of unresolved variance}

As we have noted above, the measured variance of a field observed at finite resolution is 
necessarily a lower limit to the true variance of the field.
An obvious question then arises: what about structure (variance) at scales 
below the best available resolution? 

If we assume that the observed slope of the power spectrum at high wavenumbers continues to a maximum wavenumber, $k_{c}$,
beyond which no further structure is present, we can estimate the amount of variance not included in our calculation (see BFP).
It has been suggested that $k_{c}$ may be determined by the sonic scale: turbulent motions are supersonic above this scale
and subsonic below it (V\'{a}zquez-Semadeni, Ballesteros-Paredes, \& Klessen 2003; Federrath et al 2009). Taking 
$\sigma_{v} \propto L^{0.5}$ (Heyer \& Brunt 2004), we estimate the sonic scale in Taurus as 
$L_{s} = L_{0}/M^{2} \approx$~0.08~pc where $L_{0} \approx$~25~pc is the linear size of the cloud from Figure~1, and
$M = 17.6$ is the Mach number from above. This is comparable to the spatial resolution of the data (0.03~pc),
which suggests that our measurement of $\sigma^{2}_{\rho/\rho_{0}}$ does not significantly underestimate the true
value if structure is suppressed below the sonic scale. Taking $k_{c} \longrightarrow \infty$ on the other hand
would result in a factor of $\sim$~2 underestimation of $\sigma^{2}_{\rho/\rho_{0}}$ and therefore a factor of $\sim$~$\sqrt{2}$
underestimation of $b$. Our value of $b$ must therefore be considered as a lower limit: the power spectrum
(Figure~\ref{fig:t4}) is not well measured near the estimated sonic scale ($k_{s} \sim 350$), and 
further high resolution investigations are needed to resolve this issue.

\section{Summary}

We have provided an estimate of the parameter $b$ in the proposed $\sigma^{2}_{\rho/\rho_{0}} = b^{2}M^{2}$
relation for supersonic turbulence in isothermal gas, using $^{13}$CO observations of the Taurus molecular
cloud. Using $^{13}$CO and 2MASS extinction data, 
we find $b = 0.48^{+0.15}_{-0.11}$, which is consistent with 
the value originally proposed by Padoan et al (1997) and characteristic of turbulent forcing which includes a mixture
of both solenoidal and compressive modes (Federrath et al 2009). Our value of $b$ is a lower limit if 
significant structure exists below the resolution of our observations (effectively, below the sonic scale).
If diffuse material is included in the column density budget then we find a somewhat lower value of $b = 0.32$,
although in this case there are some questions over the assumption of isothermality in the Mach number calculation.
Our most conservative statement is therefore that $b$ is constrained to lie in the range $0.3 \lesssim b \lesssim 0.8$,
which is comparable to the range of current numerical estimates. However, the $\sigma^{2}_{\rho/\rho_{0}} - M^{2}$
relation has been tested to only relatively low Mach numbers ($M \lesssim 7$) in comparison to that in
Taurus, so further numerical exploration of the high Mach number regime should be carried out. In principle,
gravitational amplification of turbulently-generated density enhancements could increase $\sigma^{2}_{\rho/\rho_{0}}$  
at roughly constant Mach number, with gravity possibly acting analogously to compressive forcing. Numerical
investigation of the role of gravity in the $\sigma^{2}_{\rho/\rho_{0}} - M^{2}$ would be  worthwhile.

Further analysis of a larger sample of molecular clouds is needed to investigate the linearity of the
$\sigma^{2}_{\rho/\rho_{0}}$~--~$M^{2}$ relation and possible variations in $b$. As our main sources
of uncertainty arise because of questions over how to treat the diffuse (likely atomic) regime and
how to account for unresolved density variance, further progress on improving the accuracy of measuring $b$
ultimately must involve answering what is meant by ``a cloud'' (Ballesteros-Paredes, V\'{a}zquez-Semadeni, \& Scalo 1999)
and down to what spatial scale is significant structure likely to be present? 

\begin{acknowledgements}
I would like to thank Dan Price and Christoph Federrath for excellent encouragement, advice,
and scientific insight.
This work was supported by STFC Grant ST/F003277/1 to the University of Exeter,
Marie Curie Re-Integration Grant MIRG-46555, and NSF grant AST 0838222 to the Five College
Radio Astronomy Observatory. CB is supported by an RCUK fellowship
at the University of Exeter, UK. The Five College Radio Astronomy Observatory was 
supported by NSF grant AST 0838222. 
This publication makes use of data products from 2MASS, which is a joint project 
of the University of Massachusetts and the Infrared Processing and Analysis 
Center/California Institute of Technology, funded by the National Aeronautics 
and Space Administration and the National Science
\end{acknowledgements}

\begin{appendix}

\section{Zero Padding}

Brunt, Federrath, \& Price (2009) considered an observable 2D field, $F_{2}$, which is produced
by {\it averaging} a 3D field, $F_{3}$, over the line-of-sight axis. The field $F_{2}$ was considered
to be distributed in a square region with a scale ratio (number of pixels along each axis) of $\lambda$.
Assuming isotropy, the 3D field then is distributed in a cubical region, again of scale ratio $\lambda$.

A column density field, $N$, is produced by {\it integrating} the density field, $\rho$, along the line-of-sight,
rather than averaging. To conform to the conditions set out by BFP, it is necessary to work with
the {\it normalised} column density field and density field, which are obtained by dividing each field
by their respective mean values, $N_{0}$ and $\rho_{0}$. The variances calculated from these fields are $\sigma^{2}_{N/N_{0}}$ 
and $\sigma^{2}_{\rho/\rho_{0}}$ respectively. The latter is the quantity required to test 
the theoretical prediction: $\sigma^{2}_{\rho/\rho_{0}} = b^{2}M^{2}$.

Consider a 2D field, $F_{2}$, with mean value $F_{2,0}$, which is the projection of a 3D field, $F_{3}$, with mean $F_{3,0}$.
The normalised variance, $\sigma^{2}_{2} = \sigma_{F_{2}/F_{2,0}}^{2}$, 
can be calculated for a square image, of scale ratio $\lambda$, via:
\begin{equation}
\sigma^{2}_{2} = \frac{\langle F_{2}^{2} \rangle - F_{2,0}^{2}}{F_{2,0}^{2}} = \frac{\langle F_{2}^{2} \rangle}{F_{2,0}^{2}} - 1
\label{eqnoa01}
\end{equation}
where:
\begin{equation}
\langle F_{2}^{2} \rangle = \frac{\displaystyle\sum_{i=1}^{\lambda} \displaystyle\sum_{j=1}^{\lambda} F_{2}^{2}(i,j)}{\lambda^{2}} 
\end{equation}
\begin{equation}
F_{2,0} = \langle F_{2} \rangle = \frac{\displaystyle\sum_{i=1}^{\lambda} \displaystyle\sum_{j=1}^{\lambda} F_{2}(i,j)}{\lambda^{2}} 
\end{equation}
and $F_{2}(i,j)$ is the value of $F_{2}$ at the pixel ($i,j$).

Similarly, the normalised variance of $F_{3}$ is $\sigma^{2}_{3} = \sigma_{F_{3}/F_{3,0}}^{2}$, given by:
\begin{equation}
\sigma^{2}_{3} = \frac{\langle F_{3}^{2} \rangle - F_{3,0}^{2}}{F_{3,0}^{2}} = \frac{\langle F_{3}^{2} \rangle}{F_{3,0}^{2}} - 1
\label{eqnoa02}
\end{equation}
where:
\begin{equation}
\langle F_{3}^{2} \rangle = \frac{\displaystyle\sum_{i=1}^{\lambda} \displaystyle\sum_{j=1}^{\lambda} \displaystyle\sum_{k=1}^{\lambda} F_{3}^{2}(i,j,k)}{\lambda^{3}}
\end{equation}
\begin{equation}
F_{3,0} = \langle F_{3} \rangle = \frac{\displaystyle\sum_{i=1}^{\lambda} \displaystyle\sum_{j=1}^{\lambda} \displaystyle\sum_{k=1}^{\lambda} F_{3}(i,j,k)}{\lambda^{3}}
\end{equation}
and $F_{3}(i,j,k)$ is the value of $F_{3}$ at the pixel ($i,j,k$).

\begin{figure}
\includegraphics[width=84mm]{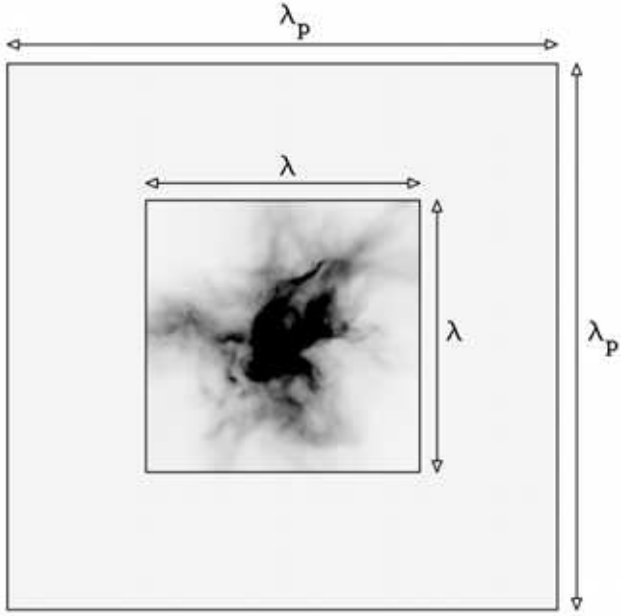}
  \caption{ Schematic illustration of zero-padding.
}
\label{fig:a1}
\end{figure}

If $F_{2}$ is now zero-padded to a scale ratio of $\lambda_{p}$ (shown schematically in Figure~\ref{fig:a1}), 
the normalised variance, $\sigma_{2p}^{2}$, of the resulting field, $F_{2p}$, is given by:
\begin{equation}
\sigma^{2}_{2p} = \frac{\langle F_{2p}^{2} \rangle - F_{2p,0}^{2}}{F_{2p,0}^{2}} = \frac{\langle F_{2p}^{2} \rangle}{F_{2p,0}^{2}} - 1
\label{eqnoa03}
\end{equation}
where now:
\begin{equation}
\langle F_{2p}^{2} \rangle = \frac{\displaystyle\sum_{i=1}^{\lambda_{p}} \displaystyle\sum_{j=1}^{\lambda_{p}} F_{2p}^{2}(i,j)}{\lambda_{p}^{2}} = \langle F_{2}^{2} \rangle \frac{\lambda^{2}}{\lambda_{p}^{2}} = \langle F_{2}^{2} \rangle / \eta^{2}
\end{equation}
\begin{equation}
F_{2p,0} = \langle F_{2p} \rangle = \frac{\displaystyle\sum_{i=1}^{\lambda_{p}} \displaystyle\sum_{j=1}^{\lambda_{p}} F_{2p}(i,j)}{\lambda_{p}^{2}} = \langle F_{2} \rangle \frac{\lambda^{2}}{\lambda_{p}^{2}} = \langle F_{2} \rangle / \eta^{2}
\end{equation}
where we have defined $\eta = \lambda_{p}/\lambda$, and noted that:
\begin{equation}
\displaystyle\sum_{i=1}^{\lambda_{p}} \displaystyle\sum_{j=1}^{\lambda_{p}} F_{2p}(i,j) = \displaystyle\sum_{i=1}^{\lambda} \displaystyle\sum_{j=1}^{\lambda} F_{2}(i,j)
\end{equation}
\begin{equation}
\displaystyle\sum_{i=1}^{\lambda_{p}} \displaystyle\sum_{j=1}^{\lambda_{p}} F^{2}_{2p}(i,j) = \displaystyle\sum_{i=1}^{\lambda} \displaystyle\sum_{j=1}^{\lambda} F^{2}_{2}(i,j)
\end{equation}
since $F_{2p}$ is zero outside the region where $F_{2}$ is defined.

Similarly:
\begin{equation}
\sigma^{2}_{3p} = \frac{\langle F_{3p}^{2} \rangle - F_{3p,0}^{2}}{F_{3p,0}^{2}} = \frac{\langle F_{3p}^{2} \rangle}{F_{3p,0}^{2}} - 1
\label{eqnoa04}
\end{equation}
\begin{equation}
\langle F_{3p}^{2} \rangle = \frac{\displaystyle\sum_{i=1}^{\lambda_{p}} \displaystyle\sum_{j=1}^{\lambda_{p}} \displaystyle\sum_{k=1}^{\lambda_{p}} F_{3p}^{2}(i,j,k)}{\lambda_{p}^{2}} = \langle F_{3}^{2} \rangle \frac{\lambda^{3}}{\lambda_{p}^{3}} = \langle F_{3}^{2} \rangle / \eta^{3}
\end{equation}
\begin{equation}
F_{3p,0} = \langle F_{3p} \rangle = \frac{\displaystyle\sum_{i=1}^{\lambda_{p}} \displaystyle\sum_{j=1}^{\lambda_{p}} \displaystyle\sum_{k=1}^{\lambda_{p}} F_{3p}(i,j)}{\lambda_{p}^{2}} = \langle F_{3} \rangle \frac{\lambda^{3}}{\lambda_{p}^{3}} = \langle F_{3} \rangle / \eta^{3} .
\end{equation}

Combining these results, we find:
\begin{equation}
\sigma^{2}_{2p} = (\sigma_{2}^{2} + 1)\eta^{2} - 1
\label{eqnoa1}
\end{equation}
\begin{equation}
\sigma^{2}_{3p} = (\sigma_{3}^{2} + 1)\eta^{3} - 1 .
\label{eqnoa2}
\end{equation}

Application of the BFP method to the zero-padded field yields the ratio of 2D-to-3D normalised variance:
\begin{equation}
R_{p} = \sigma^{2}_{2p} / \sigma^{2}_{3p}
\label{eqnoa3}
\end{equation}
where $R_{p}$ is calculated using the power spectrum of the zero-padded field, and we have identified
this by the subscript $p$ on $R$. The quantity of interest, however, is $\sigma_{3}^{2}$ which can be
derived from the measured $\sigma_{2}^{2}$ and $R_{p}$ via:
\begin{equation}
\sigma^{2}_{3} = \frac{1}{\eta^{3}} ( 1 + ((\sigma^{2}_{2} + 1)\eta^{2} -1 )/R_{p}) - 1 ,
\label{eqnoa4}
\end{equation}
obtained through combining equations~\ref{eqnoa1}, \ref{eqnoa2}, and \ref{eqnoa3}. 

The BFP method assumes that the input image from which the 2D power spectrum is calculated is
square. In the case that the observed field is not square, but has dimensions $\lambda_{x} \times \lambda_{y}$,
zero-padding to a square field of size $\lambda_{px} \times \lambda_{py}$ is required. If now $\eta$ is defined as:
\begin{equation}
\eta = \sqrt{\lambda_{px}\lambda_{py}/\lambda_{x}\lambda_{y}} ,
\end{equation}
this allows $\sigma_{3}^{2}$ to be calculated via equation~\ref{eqnoa4} under the assumption that
the line-of-sight extent of the field $F_{3}$ is:
\begin{equation}
\lambda_{z} = \sqrt{\lambda_{x}\lambda_{y}}
\end{equation}
Obviously, it is desirable that $\lambda_{x} \approx \lambda_{y}$ for the assumption of isotropy to
be best satisfied.

\end{appendix}

\end{document}